# Pedestrian Static Trajectory Analysis of a Hypermarket

Kardi TEKNOMO
Associate Professor
Information Systems and Computer Science
Ateneo de Manila University
Loyola Heights, Quezon City
Philippines
Fax: +63-2-426-6071
E-mail: teknomo@gmail.com

Gloria P. GERILLA
Assistant Professor
Department of Civil Engineering
De LaSalle University
Taft Avenue, Manila
Philippines
Fax: +63-2-635-2693
E-mail: glogtek@gmail.com

**Abstract:** In this paper, we propose a combination of pedestrian data collection and analysis and modeling that may yield higher competitive advantage in the business environment. The data collection is only based on simple inventory and questionnaire surveys on a hypermarket to obtain trajectory path of pedestrian movement. Though the data has limitation by using static trajectories, our techniques showed that it is possible to obtain aggregation of flow pattern and alley attractiveness similar to the result of aggregation using dynamic trajectory. A case study of a real hypermarket demonstrates that daily necessity products are closely related to higher flow pattern. Using the proposed method, we are also able to quantify pedestrian behavior that shoppers tend to walk about 7 times higher than the ideal shortest path.

## 1. INTRODUCTION

With the growth of a city into larger regions and new urban city centers are repositioned, traditional markets have been transformed gradually into groceries stores, supermarkets, and hypermarket. As business becomes more competitive, new strategies to improve shopping experience as well as competing in price and product and service quality and lifestyle are necessary. For a shopping center to stay with competitive advantage, it needs better knowledge on shopping behavior and more accurate measurement on shoppers (i.e. pedestrian) behavior.

While traditional demand on the shopping center are derived from counting the number of people coming in the shopping center, new types of gadgets allow shopping centers' owners to track each pedestrian traveling around the shops over time and space. The result of such gadget is a *pedestrian trajectory* which is an abstraction of path taken by a pedestrian in a system from the time a pedestrian enters the system until pedestrian goes out of the system. Dynamic pedestrian trajectory can be obtained through pedestrian tracking technology such as such as GPS, Wireless (mobile phone and PDA) as well as RFID and multiple view of video camera are used for both indoor and outdoor business environment (Teknomo 2006). Such trajectory contains dynamic data along the path that the shoppers passed in the mall. The aggregation of dynamic trajectory analysis over time produces *flow pattern* which represents attractiveness of alleys based on the frequency of trajectory in the alley.

For many practical applications, especially in developing countries where the labor cost is much cheaper than electronic devices, however, such pedestrian tracking technologies and gadgets are still very much costly and require advanced knowledge on its utilization. A more straightforward technique is needed to gain flow pattern resulting from the trajectory analysis. In this paper, we propose a simple technique to obtain trajectory data merely using questionnaire survey. Compared to advanced pedestrian tracking technology, our technique has limitation to obtain only static data. Thus, we give name to such pedestrian path as *static trajectory*. However, in this paper we will show that such limitation does not hinder us to obtain similar results as dynamic trajectory analysis in terms of flow pattern and measurement of alley attractiveness.

Using static trajectory analysis, the objective of this paper is to answer several important questions such as which alleys on the stores get more flow than the others and which product-items are preferable than the others. To be precise, our goal in trajectory analysis is to find flow pattern and finally to derive alley attractiveness. To the best of our knowledge, this is the first paper to describe about static trajectory from questionnaire survey to analyze pedestrian and shopper's behavior.

The future applications of static trajectory analysis, however, are much farther beyond what is addressed in this paper. For example, the static trajectory analysis has potential applications like the following:

- **Evaluation of marketing strategy**: Panel data before and after marketing strategy (e.g. discount, promotion, bargain sales, etc.) could show how well the marketing strategy is.
- **Understanding shopper behavior**: Knowing pedestrian flow data is also useful to learn the behavior of shoppers. Weekend visitors, for example, have different shopping behavior characteristics than weekday visitors, lunch time visitors mostly coming for food rather than to buy clothes. Pedestrian flow data over time explain the cyclical demand and trend of number of visitors. Business needs to know when the peak and lowest number of visitors is such that all the marketing strategy can be targeted to more appropriate customers.
- **Provide best design for best rental prices**: to make almost all alleys have pedestrian flow beyond a minimum threshold level. Bad design if certain alleys receives only very few pedestrian flow. Flow pattern may also reveal relative shop prices (or rental fee) based on alley attractiveness. Higher flow pattern goes through the alley must be higher demand, therefore higher price per floor area.
- **Finding best shop location**: for shop owner who would like to find a place where the pedestrian flow is the highest (but still under certain threshold of crowd level) or within some specified preferable range
- **Best design of shop layout**: Trajectory analysis (from both real data and pedestrian simulation) may reveal preference map: which alleys on the stores get more flow than the others.

The remainder of the paper is organized as follows. The next section, we describe how the data collection was done based on inventory survey and questionnaire survey, then network analysis of inventory survey is described in Section 3. In Section 4 and 5, we explain the analysis of trajectory based on combination of the questionnaire survey and inventory survey. Section 4 describes the behavioral inference of shoppers' walking distance while Section 5 emphasizes the flow pattern and alley attractiveness indices that are important for business application. The analysis of the two surveys is combined in Section 5. Finally conclusions are drawn in the last section.

## 2. DATA COLLECTION

To proof our concept on static trajectory analysis, we utilize the data that have been gathered by Siahaan in a hypermarket in Surabaya as mere case study. This section explains only the data collection process in relation to our concept. Readers who are interested to read further about the questionnaire data can refer to (Siahaan, 2007). Two types of data collection were done. The first data collection is an inventory of the hypermarket layout configuration and products sellable along all the alleys by category. The second data collection was done using questionnaire survey.

The inventory data collection was performed with the aid of an architectural layout sketch of the hypermarket where the geometrical measurement has been posted. Adjustment of the geometry (length and width in top view) was observed and noted as the surveyor walks along the alleys. Figure 1 illustrates the results of inventory survey. The hypermarket is divided into several sections such as groceries, clothing, electronics, home furnishing and so on. There is only a single main entry which is also utilized as an exit point and there is a single main exit. There are also several peripheral shops located between the two main gates. For our study, the main focus is only pedestrian behavior inside the hypermarket and ignores the trajectories toward the peripheral shops.

Figure 1. Results of the inventory survey

To quantify our proposed trajectory analysis above, a questionnaire survey with in depth interview was done in Giant Hypermarket Surabaya, Indonesia. The questions in the survey consist of two main parts:
1. Trajectory of each shopper in the hypermarket
2. Shopping behavior which consists of socio economic data (gender, net income, education level, type of work and age), activities in the mall (such as shopping purpose, duration, number of companions and total spending) and shopping trip characteristics such as shopping frequency.

For the scope of this paper, however, our focus will be on the analysis of first part of the questionnaire survey. The statistical analysis of the second parts of the questionnaire is omitted because it has been reported elsewhere (Siahaan, 2007).

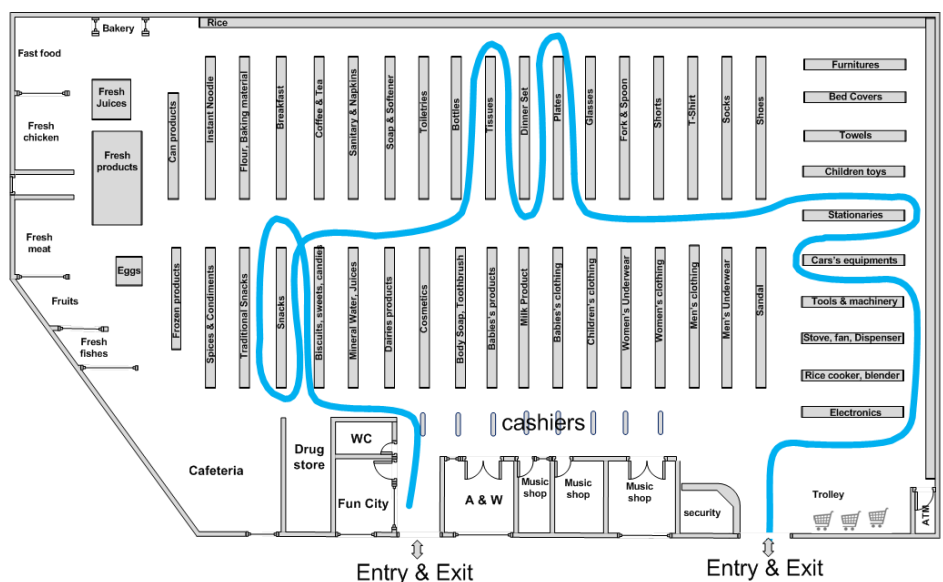

Figure 2. Example questionnaires and a static trajectory drawn by a shopper #160

To obtain pedestrian static trajectory via questionnaire survey, each respondent were requested to draw their route on provided floor plan or map of the shops. Figure 1 shows the example of a trajectory path drawn by a respondent. The interview was done after the respondents went out of the cashiers, thus they have completed their shopping journey inside the hypermarket. We call the trajectory *static trajectory* because the time information is missing.

At preliminary survey, time stamp data and shopping items that they picked along the route were also included. The attempt to include time stamp along the path was done with much difficulties as most respondents have forgotten it. We also found out that the shopping items data along the route were not so accurate compared to their actual list of shopping items, therefore we disregard these data and retain only the route or static trajectories because of its high accuracy.

Each interview took about 10 to 20 minutes during weekdays. As the time is not part of equation, the survey can be done even during off peak. Total respondents were 204 with only 188 respondents considered as complete and valid. The invalid data were mostly due to incomplete trajectories or their trajectories are not readable.

## 2. NETWORK ANALYSIS

From the inventory survey and architectural layout plan, we select a position in the middle of the alleys to be the nodes of the network graph while each alley is represented by a link. Figure 3 illustrates the network graph of walkway along the hypermarket. The coding of the nodes is arbitrary but we prefer to be represented in alphabetical order from A-Z, AA-BZ and 1a-1j. In total we employ 88 nodes and 139 links. Some link such as BX-BZ consists of three actual nodes of the graph representation but for the analysis we only employ 2 nodes for simplicity without losing generality. Except the links that pass through the cashiers which is only one way, all the other links are two ways because the pedestrians can walk in two directions.

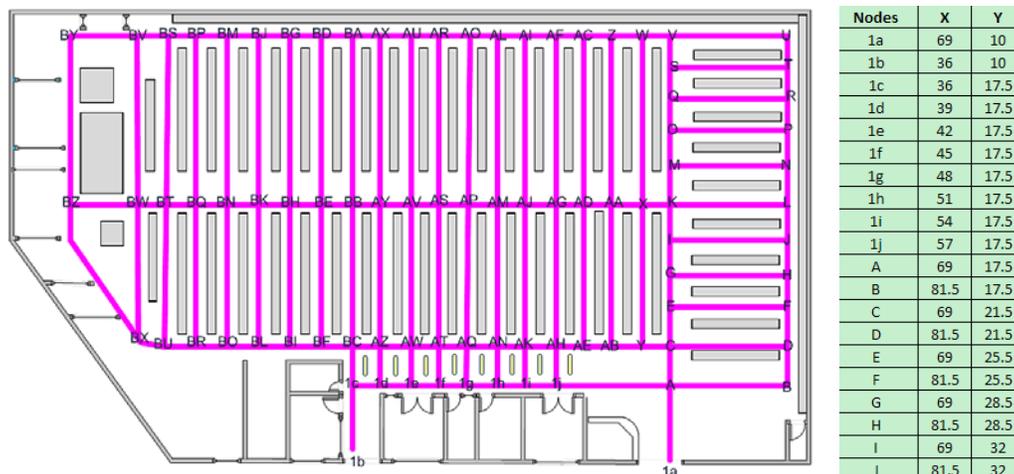

Figure 3. Graph model of alleys in the hypermarket and sample of nodes coordinates

The coordinates of each node were measured directly from the map (see example on Figure 3 right) and adjacency matrix **A** to represent the structure of the network was derived. Distance matrix **D** indicates the length of each link was computed based on Euclidean distance between coordinates of two nodes. The value of each element in a distance matrix represents length of the link (connected nodes). Unconnected vertices are represented by infinity distance.

The distance matrix and the adjacency matrices are sparse matrix with most of the entries located near the diagonal elements but the diagonal elements themselves are zero. As the matrix size is rather large (88 by 88), it is valuable to scale the matrix values into a color map of an image to see the pattern. The left hand side of Figure 4 shows the distance matrix that is mapped into an image with a color scale that indicates the brighter color with a higher value. Unsymmetrical on the middle left of the image is due to one way links passing through the cashiers. The actual values of the top-left part of the distance matrix are shown on the right of Figure 4 together with the coding of the nodes.

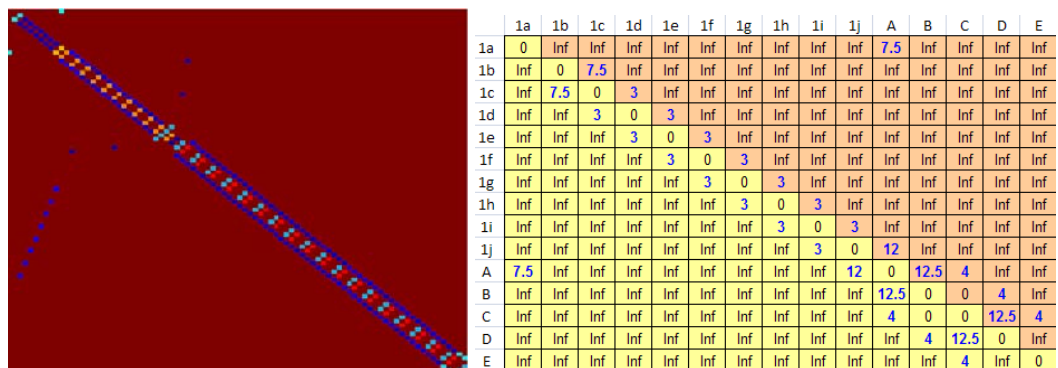

Figure 4. Mapping of distance matrix (left) and top left corner of the distance matrix (right)

The geometric arrangement of the alleys inside a hypermarket can be measured indirectly through statistical indices of the distribution of shortest path distances between all pairs of nodes. The shortest path distance was computed using the Floyd-Warshall algorithm (Cormen et al, 2001), which finds shortest paths between all pairs of nodes in the network graph. Figure 5 left shows the mapping of all-pairs shortest-path distance matrix into heating color map. Higher distance has heating red color and lower distance value is indicated by a cool blue color. The figure clearly shows that the shortest path distance matrix values are not only on the surrounding diagonal elements as the distance matrix. The diagonal elements are all zeros. The value of each element in the shortest distance matrix represents shortest path length between any two nodes on the network graph. The right Figure 5 illustrates the top left values of the all pair shortest path distance matrix. Comparing these values to the distance matrix in Figure 4 right, we can see that the non-infinity values of the distance matrix are actually embedded inside the shortest path distance matrix.

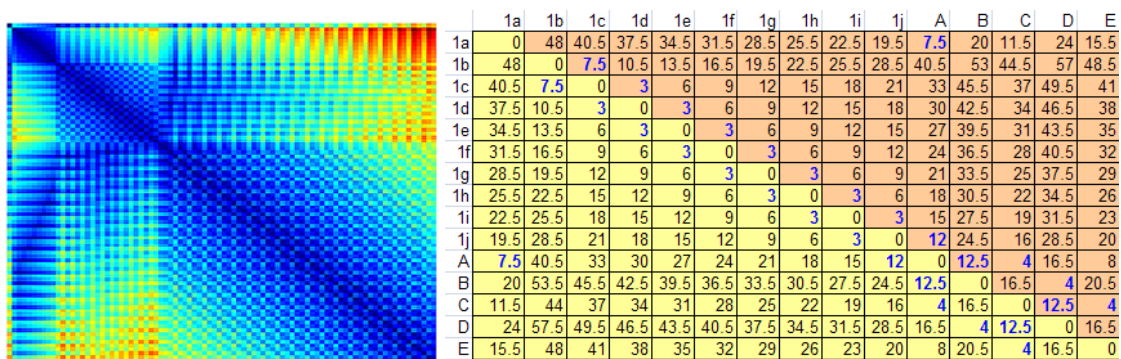

Figure 5. Mapping of all shortest path distance matrix and top left corner of the matrix

The distribution of all non-zero pairs with the shortest path distance is skewed to the right as shown in Figure 6. The mean of the distribution is 38.87 meters (95% confidence interval is between 38.87 to 38.88 meters) while the median is 36.5 meters and mode of 27.5. The maximum shortest path distance can reach up to 123.1 meters with standard deviation of 21.4 meters and inter quartile range of 30.5 meters. The Box plot on the bottom of Figure 6 indicates skewness of 0.5573 and distance above 98.75 meters can be considered as outliers. We will compare these figures later with the results of trajectories analysis in Section 4.

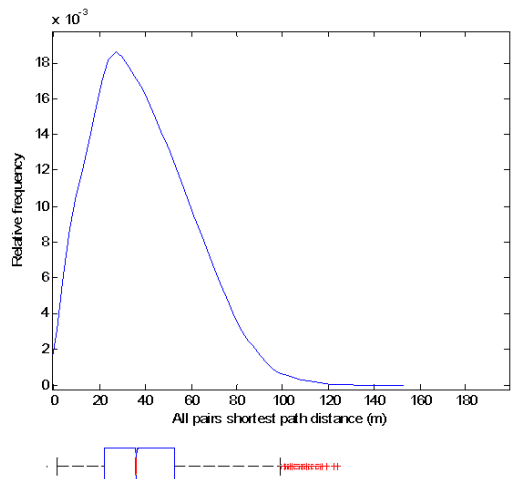

Figure 6. Distribution of all pairs shortest path distance matrix and its box plot

**4. STATIC TRAJECTORY ANALYSIS**

In this and the next sections we describe the analysis results of the questionnaire survey on trajectory path of the shopper and combine them with the results of inventory survey that has been explained in the previous section.

Pedestrian trajectory can be represented as an NTXY database (first proposed by Teknomo et al 2000) which consists of a pedestrian ID number, time slice when the data is recorded from the top view 2 dimensional coordinate based on some local reference coordinate system. Notation N indicates pedestrian number, T represents time stamp where the trajectory are recorded along each coordinate (X, Y). The tracking is performed for each pedestrian over time and space. A static trajectory is a reduction of pedestrian trajectory to include only space list of coordinates for each pedestrian (i.e. only NXY).

From the questionnaire survey, pedestrian trajectory is drawn by the respondent to indicate his or her journey inside the hypermarket. The trajectory drawing is then mapped into the nearest links and nodes of the network graph and coded as node sequence. For example, for a shopper with code number 160 where its trajectory was drawn in Figure 2, has nodes sequence of 1a-A-B-D-F-H-J-I-K-L-N-M-K-X-AA-AD-AG-AJ-AM-AL-AO-AP-AS-AR-AU-AV-AY-BB-BE-BH-BK-BL-BO-BN-BQ-BR-BO-BL-BI-BF-BC-AZ-1d-1c-1b as illustrated in Figure 7.

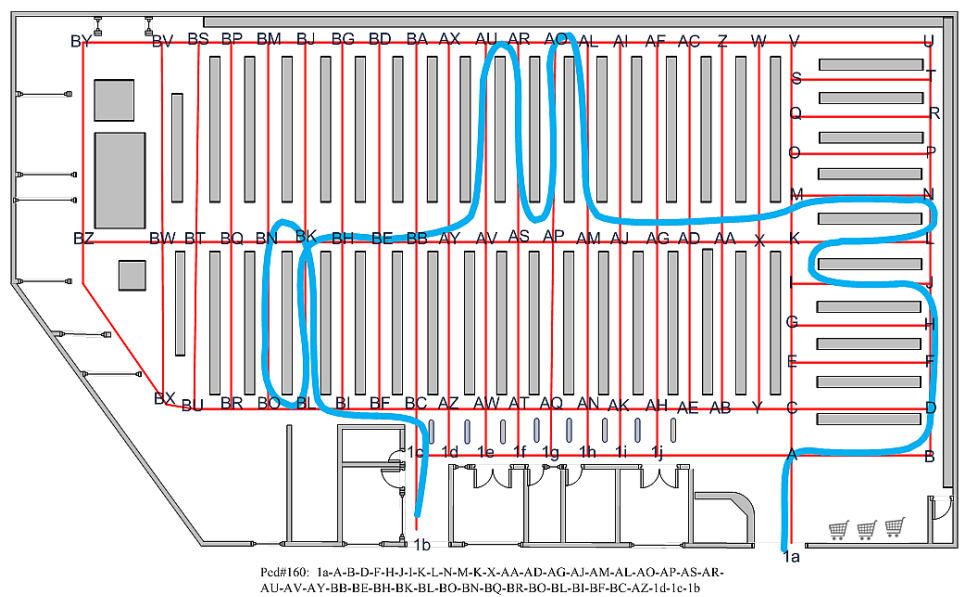

Figure 7. Mapping of trajectory of a shopper into network graph produces nodes sequence.

Once the mapping of the trajectories drawn by the respondents was done for all 188 samples, our static trajectories become a set of node sequences. As the coordinates of each node is known, we can easily compute the walking distance of each pedestrian inside the hypermarket from the node sequence as cumulative of length of the links along the trajectory path of the shopper. We must give note, however, that the walking distance computed is not the actual walking distance but only approximation of the walking distance because the real pedestrian trajectories has been transformed twice from actual behavior into the respondent's drawing pad and then mapped into the network graph.

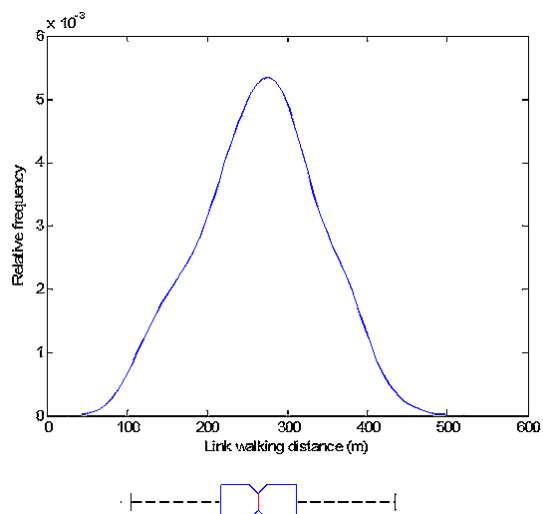

Figure 8. Distribution and box plot of link walking distance

Figure 8 exhibits the distribution of link walking distance for all 188 shoppers in the sample. The mean of the distribution of walking distance is 265.14 meters (95% confidence interval is between 264.40 to 265.88) while the median is 267.8 meters and mode of 152 meters. The maximum link-walking distance is about 430 meters with standard deviation of 70.82 meters and interquartile range of 95 meters.

The Box plot on the bottom of Figure 8 indicates skewness of -0.15 close to symmetric distribution and walking distance below 77.8 meters or above 457.8 meters can be considered as outliers. It would be remarkable to compare this outlier boundary value to the central tendency of all the pairs with shortest path distance because more than 95% of the distribution of shortest path distance is under the lower bound outlier threshold value of the link walking distance.

Based on the above finding, we propose a simple index to compare the distribution of the walking distance and the distribution of non-zero all pairs shortest path distance. Our indicator is simply a median ratio

$$\lambda = \frac{\text{median link walking distance}}{\text{median all pair shortest path distance}} \quad (1)$$

We use median instead of mean because the skewness of the two distributions are not equal and median is more robust central tendency than mean. The value of our indicator of median ratio $\lambda$ is always higher than one because we compare the aggregation of actual walking distance to the ideal shortest path walking distance for all cases. Nearer the value of $\lambda$ to unity may indicate simplicity of the layout arrangement that the shoppers can go straightforward almost similar to the shortest path from any points.

In our case study, we have $\lambda = 267.8/36.5 = 7.34$. It means that the actual walking distance is 7 times longer than the ideal shortest path. In other words, shoppers do not walk in the shortest path which is probably a well known common sense. However, our indicator and our case study are able to yield a quantitative verification of this common sense.

## 5. FLOW PATTERN AND ALLEY ATTRACTIVENESS

The main objective of trajectory analysis is to find flow pattern and finally to derive alley attractiveness. In this section, we will describe how we will obtain the alley attractiveness from static trajectory analysis. The alley attractiveness is useful to answer several important business questions such as which alleys on the stores get more flow than the others and which product-items are preferable than the others. First, we will clarify the terminologies and assumptions, and then we will detail the computation and results of the analysis.

Using static trajectory analysis, the frequency of trajectories passing through each alley is aggregated to represent *flow pattern*. We called it 'flow' rather than 'density' because they come from frequency aggregation of trajectories over a time period of the study rather than within an instant of time. Some critical readers may argue that the static trajectory come from questionnaire samples which do not represent true flow or density. However, flow pattern is not equal to the flow rate in traffic engineering sense. Flow pattern is an index based on flow to show which alley has higher visitors compared to the other alleys. To distinguish from the term 'flow rate' that is normally used in traffic engineering studies; we give a name 'flow pattern' because it does indeed represent the pattern of the flow over different alleys in a business environment.

The flow pattern on each alley represents demand to that particular alley as the people pass through the alley and see the display or the product items, there is a higher chance that the product items would be picked up and put into their shopping baskets. For the designer or owner of shopping center and business environment, flow patterns give very useful information on which part of stores have higher flow than the others parts. Higher pedestrian flow pattern is highly correlated to the rental fee of the stores.

When the flow pattern is derived from dynamic trajectories (that can be obtained through tracking devices such as RFID and GPS), the trajectories are aggregated as frequency of passing the alley over small time interval (e.g. 5 minutes or 15 minutes) and finally aggregated again by averaging them over the time interval of the study period or over the number of the flow. This final aggregation is basically static that it does not contain any information regarding time (except the study period). Assuming the same amount of total time study period, the aggregation of static trajectories over all samples (in the limit sense that we have infinitely many of such static trajectories) would eventually approach the equivalent of the final aggregation of the dynamic trajectories. Hence, we can utilize the information of the static trajectories as an approximation of the real flow pattern.

The computation of flow pattern is straightforward as counting the frequency of trajectories passing through each alley. An alley is represented by a link in the network graph. As the trajectories have been transformed into nodes sequences and, the link frequency can be put into a flow matrix **F** which is equivalent to the adjacency matrix **A** with all ones elements being replaced with the flow pattern. Table 1 shows the pseudo code for the algorithm to compute flow matrix. The flow matrix contains the flow pattern or frequency of each link on the network graph on each direction over the study time period.

Table 1: Algorithm for computation of flow pattern

```
For each respondent
    For each trajectory
        For each link in the trajectory
            F (node origin, node destination) = F (node origin, node destination) + 1
        Next link
    Next trajectory
Next respondent
```

The values of flow matrix **F** are illustrated in Figure 9. The flow matrix has a similar pattern to distance matrix (Figure 4) but the values are different. Higher flow has heat orange color and lower distance value is indicated by a cool blue color. The maximum value of flow pattern is 265 pedestrians per alley.

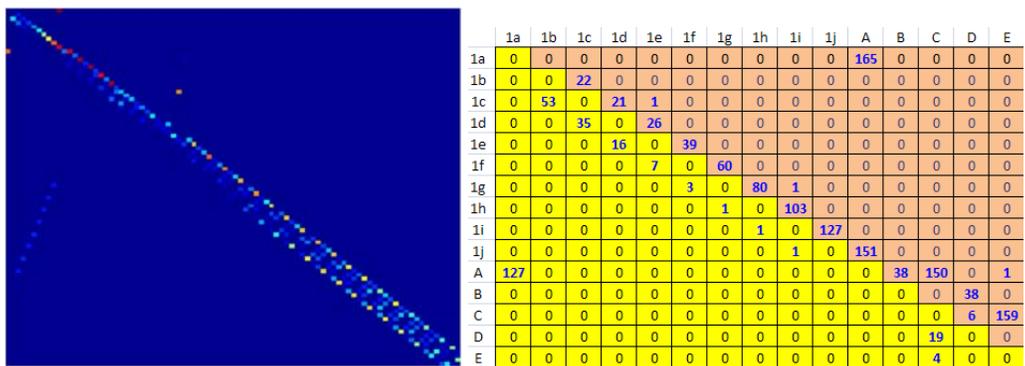

|    | 1a  | 1b | 1c | 1d | 1e | 1f | 1g | 1h | 1i  | 1j  | A   | B  | C   | D  | E   |
|----|-----|----|----|----|----|----|----|----|-----|-----|-----|----|-----|----|-----|
| 1a | 0   | 0  | 0  | 0  | 0  | 0  | 0  | 0  | 0   | 0   | 165 | 0  | 0   | 0  | 0   |
| 1b | 0   | 0  | 22 | 0  | 0  | 0  | 0  | 0  | 0   | 0   | 0   | 0  | 0   | 0  | 0   |
| 1c | 0   | 53 | 0  | 21 | 1  | 0  | 0  | 0  | 0   | 0   | 0   | 0  | 0   | 0  | 0   |
| 1d | 0   | 0  | 35 | 0  | 26 | 0  | 0  | 0  | 0   | 0   | 0   | 0  | 0   | 0  | 0   |
| 1e | 0   | 0  | 0  | 16 | 0  | 39 | 0  | 0  | 0   | 0   | 0   | 0  | 0   | 0  | 0   |
| 1f | 0   | 0  | 0  | 0  | 7  | 0  | 60 | 0  | 0   | 0   | 0   | 0  | 0   | 0  | 0   |
| 1g | 0   | 0  | 0  | 0  | 0  | 3  | 0  | 80 | 1   | 0   | 0   | 0  | 0   | 0  | 0   |
| 1h | 0   | 0  | 0  | 0  | 0  | 0  | 1  | 0  | 103 | 0   | 0   | 0  | 0   | 0  | 0   |
| 1i | 0   | 0  | 0  | 0  | 0  | 0  | 0  | 1  | 0   | 127 | 0   | 0  | 0   | 0  | 0   |
| 1j | 0   | 0  | 0  | 0  | 0  | 0  | 0  | 0  | 1   | 0   | 151 | 0  | 0   | 0  | 0   |
| A  | 127 | 0  | 0  | 0  | 0  | 0  | 0  | 0  | 0   | 0   | 0   | 38 | 150 | 0  | 1   |
| B  | 0   | 0  | 0  | 0  | 0  | 0  | 0  | 0  | 0   | 0   | 0   | 0  | 0   | 38 | 0   |
| C  | 0   | 0  | 0  | 0  | 0  | 0  | 0  | 0  | 0   | 0   | 0   | 0  | 0   | 6  | 159 |
| D  | 0   | 0  | 0  | 0  | 0  | 0  | 0  | 0  | 0   | 0   | 0   | 0  | 19  | 0  | 0   |
| E  | 0   | 0  | 0  | 0  | 0  | 0  | 0  | 0  | 0   | 0   | 0   | 4  | 0   | 0  | 0   |

Figure 9. Pattern of flow matrix (left) and its top right values (right)

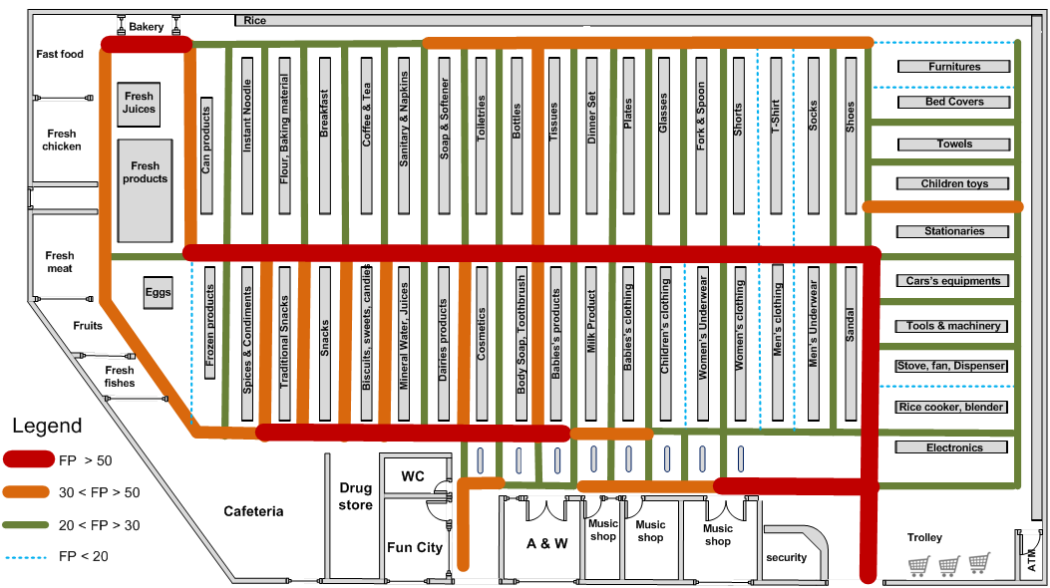

Figure 10. Spatial distribution of flow pattern

To see the spatial distribution of flow pattern (in term of total trajectories passing each alley), the values of the flow patterns are divided into 4 categories to indicate very high, high, medium and low values such that they can be scaled into the thickness of the network graph as shown in Figure 10. This spatial distribution information is very useful to distinguish which alleys on the hypermarket have more flow than the others. Relating flow pattern with the products and services along these alleys reveals the relationship between the design of store layout and products arrangement. Changing the store layout and rearranging goods inside the store will eventually change the trajectory movement pattern of visitors to move around and in turn alter the spatial pattern of the indices.

In our case study, we attempted to correlate the spatial distribution of flow pattern with the product items on the hypermarket. Relating the products along the high alley attractiveness index of Figure 10 points out that high flow pattern is related to two factors. The first major factor is that they are located in the main major route to the items that the shoppers actually want to buy. The second factor that signifies high flow pattern is the alley containing product items of daily usage such as food, cosmetics and groceries. It is interesting to see the products that related to low alley attractiveness index (dash line in Figure 10): rice cooker, mixer, stove, fan, dispenser, furniture, bed cover, adult's clothing and frozen products. These product items are clearly non-daily usage and therefore less attractive than the other necessity products.

Though flow pattern indicates the demand of each alley, the value of flow pattern itself is affected by the number of samples. As the number of sample is affected by the study area and time, the distribution of flow pattern values would be differ. This is because flow pattern gives only absolute values for the particular alley in the study area. When we would like to compare the different supermarket or hypermarket or different study period on the same study area, we need only relative values of flow pattern.

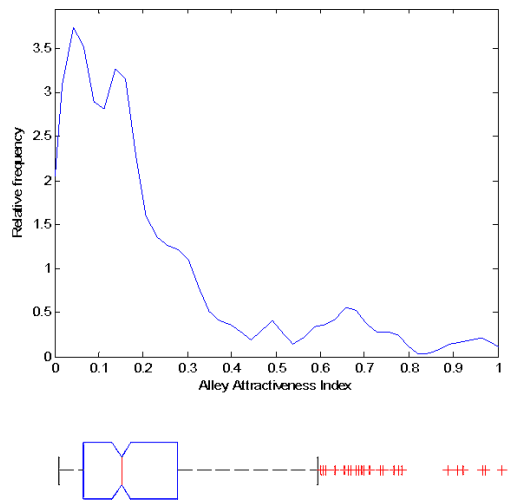

Figure 11. Distribution and box plot of alley attractiveness index values

To obtain relative values of flow pattern, we need to normalize the values of flow pattern (by division toward maximum flow pattern) to make it into an index between 0 and 1. We call the normalization of flow pattern as Alley Attractiveness Index denoted by $\alpha$ because it represents the attractiveness of each alley. Matrix of alley attractiveness index **α** as formulated in Equation (2) has exactly the same pattern as flow pattern (see Figure 9) as it is only scaled by a scalar of maximum flow pattern.

$$\boldsymbol{\alpha} = \frac{\mathbf{F}}{\max(\mathbf{F})} \tag{2}$$

Alley Attractiveness Index shows which alley inside the hypermarket has higher visitors compared to the others. The distribution of alley attractiveness indices as shown in Figure 11 is clearly skewed to the right with skewness index 1.65. The mean distribution of $\alpha$ is 0.2147 (95% confidence interval is between 0.2130 and 0.2163) with median 0.1455 and mode of 0.0061. The standard deviation is 0.2223 and interquartile range of 0.2121. In our case study, the values of $\alpha > 0.59 \approx 0.6$ are considered outliers. As the value of alley attractiveness index is between 0 and 1, one can easily multiply that by the maximum rent value to obtain the relative rental values of shops along the alley.

## 6. CONCLUSIONS

We have shown a simple yet novel method to collect static trajectory data from questionnaire and analyze them into shopper's behaviors and flow pattern or alley attractiveness indices. The alley attractiveness has been demonstrated through case study that it is useful to answer important business questions such as which alleys on the stores get more flow than the others and which product-items are preferable than the others. For our case study, daily necessity products are closely related to higher flow pattern.

We also found out from our case study that the actual walking distance is about 7 times longer than the ideal shortest path. Thus, significantly we can state that shoppers do not walk in shortest path. Shopping walking distance inside the hypermarket below 77.8 meter or above 457.8 meters can be considered as unusual or outliers.

## ACKNOWLEDGEMENTS

The authors are grateful for the questionnaire data collected by Andi Siahaan and Harry Patmadjaja from Petra Christian University, Surabaya Indonesia and also for the support of the management team from Giant Hypermarket Pondok Chandra Surabaya.